\documentclass[twocolumn, superscriptaddress, amsfonts, amssymb, amsmath]{revtex4-2}

\usepackage{graphicx, subfiles, enumitem}
\usepackage[colorlinks=true, allcolors=blue]{hyperref}
\usepackage{nameref}
\usepackage{siunitx}
\usepackage{physics}
\usepackage{subcaption}
\usepackage{makecell} 
\usepackage{float}
\usepackage[nolist]{acronym}           
\usepackage[labelfont=bf, font=small, labelsep=period, justification=raggedright]{caption}

\setcitestyle{square}
\AtBeginDocument{\RenewCommandCopy\qty\SI} 

\begin{document}

\title{A Memristive Neural Decoder for\\ Cryogenic Fault-Tolerant Quantum Error Correction}


\author{Victor~Yon}
\thanks{These authors contributed equally to this work.}
\affiliation{Irr\'eversible Inc., Sherbrooke, Qu\'ebec, Canada}
\affiliation{Institut Interdisciplinaire d’Innovation Technologique (3IT),
Universit\'e de Sherbrooke, Sherbrooke, Qu\'ebec, Canada}
\affiliation{Laboratoire Nanotechnologies Nanosyst\`emes (LN2 - IRL 3463) -- CNRS, Universit\'e de Sherbrooke, Sherbrooke, Qu\'ebec, Canada}
\affiliation{Institut quantique (IQ), Universit\'e de Sherbrooke, Sherbrooke, Qu\'ebec, Canada}

\author{Fr\'ed\'eric~Marcotte}
\thanks{These authors contributed equally to this work.}
\affiliation{1QB Information Technologies (1QBit), Vancouver, British Columbia, Canada}
\affiliation{Institut Interdisciplinaire d’Innovation Technologique (3IT),
Universit\'e de Sherbrooke, Sherbrooke, Qu\'ebec, Canada}
\affiliation{Laboratoire Nanotechnologies Nanosyst\`emes (LN2 - IRL 3463) -- CNRS, Universit\'e de Sherbrooke, Sherbrooke, Qu\'ebec, Canada}
\affiliation{Institut quantique (IQ), Universit\'e de Sherbrooke, Sherbrooke, Qu\'ebec, Canada}

\author{Pierre-Antoine~Mouny}
\affiliation{Irr\'eversible Inc., Sherbrooke, Qu\'ebec, Canada}
\affiliation{Institut Interdisciplinaire d’Innovation Technologique (3IT),
Universit\'e de Sherbrooke, Sherbrooke, Qu\'ebec, Canada}
\affiliation{Laboratoire Nanotechnologies Nanosyst\`emes (LN2 - IRL 3463) -- CNRS, Universit\'e de Sherbrooke, Sherbrooke, Qu\'ebec, Canada}
\affiliation{Institut quantique (IQ), Universit\'e de Sherbrooke, Sherbrooke, Qu\'ebec, Canada}

\author{Gebremedhin~A.~Dagnew}
\affiliation{1QB Information Technologies (1QBit), Vancouver, British Columbia, Canada}

\author{Bohdan~Kulchytskyy}
\affiliation{1QB Information Technologies (1QBit), Vancouver, British Columbia, Canada}

\author{Sophie~Rochette}
\affiliation{Irr\'eversible Inc., Sherbrooke, Qu\'ebec, Canada}

\author{Yann~Beilliard}
\affiliation{1QB Information Technologies (1QBit), Vancouver, British Columbia, Canada}
\affiliation{Institut Interdisciplinaire d’Innovation Technologique (3IT),
Universit\'e de Sherbrooke, Sherbrooke, Qu\'ebec, Canada}
\affiliation{Laboratoire Nanotechnologies Nanosyst\`emes (LN2 - IRL 3463) -- CNRS, Universit\'e de Sherbrooke, Sherbrooke, Qu\'ebec, Canada}
\affiliation{Institut quantique (IQ), Universit\'e de Sherbrooke, Sherbrooke, Qu\'ebec, Canada}

\author{Dominique~Drouin}
\affiliation{Irr\'eversible Inc., Sherbrooke, Qu\'ebec, Canada}
\affiliation{Institut Interdisciplinaire d’Innovation Technologique (3IT),
Universit\'e de Sherbrooke, Sherbrooke, Qu\'ebec, Canada}
\affiliation{Laboratoire Nanotechnologies Nanosyst\`emes (LN2 - IRL 3463) -- CNRS, Universit\'e de Sherbrooke, Sherbrooke, Qu\'ebec, Canada}
\affiliation{Institut quantique (IQ), Universit\'e de Sherbrooke, Sherbrooke, Qu\'ebec, Canada}

\author{Pooya~Ronagh}
\email[Corresponding author: ]{pooya@irreversible.tech}
\affiliation{Irr\'eversible Inc., Sherbrooke, Qu\'ebec, Canada}
\affiliation{Institute for Quantum Computing, University of Waterloo, Waterloo, Ontario, Canada}
\affiliation{Department of Physics \& Astronomy, University of Waterloo, Waterloo, Ontario, Canada}
\affiliation{Perimeter Institute for Theoretical Physics, Waterloo, Ontario, Canada}


\begin{abstract} Neural decoders for \ac{QEC} rely on
 neural networks to classify syndromes extracted from error correction codes
 and find appropriate recovery operators to protect logical information against
 errors. Its ability to adapt to hardware noise and long-term
 drifts make neural decoders a promising
 candidate for inclusion in a fault-tolerant quantum architecture. However,
 given their limited scalability, it is prudent that small-scale (local) neural
 decoders are treated as first stages of multi-stage decoding schemes for
 fault-tolerant quantum computers with millions of qubits. In this case,
 minimizing the decoding time to match the stabilization measurements frequency
 and a tight co-integration with the QPUs is highly desired. Cryogenic
 realizations of neural decoders can not only improve the performance of higher
 stage decoders, but they can minimize communication delays, and alleviate
 wiring bottlenecks. In this work, we design and analyze a neural decoder
 based on an \ac{IMC} architecture, where crossbar arrays of
 resistive memory devices are employed to both store the synaptic weights of
 the neural decoder and perform analog matrix--vector multiplications. In
 simulations supported by experimental measurements, we investigate the impact
 of TiO$_\textrm{x}$-based memristive devices' non-idealities on decoding
 fidelity. We develop hardware-aware re-training methods to mitigate the
 fidelity loss, restoring the ideal decoder's pseudo-threshold for the
 distance-3 surface code. This work provides a pathway to scalable, fast,
 and low-power cryogenic IMC hardware for integrated fault-tolerant \ac{QEC}.
\end{abstract}

\acresetall 

\date{\today}
\maketitle


\begin{acronym}
    \acro{ADC}{analog-to-digital converter}
    \acroindefinite{ADC}{an}{a}
    \acro{ASIC}{application-specific integrated circuit}
    \acroindefinite{ASIC}{an}{a}
    \acro{CMOS}{complementary metal-oxide-semiconductor}
    \acro{DAC}{digital-to-analog converter}
    \acro{DS}{device-specific}
    \acro{EV}{evaluation module}
    \acro{FPGA}{field-programmable gate array}
    \acroindefinite{FPGA}{an}{a}
    \acro{FTQC}{fault-tolerant quantum computation}
    \acroindefinite{FTQC}{an}{a}
    \acro{HCS}{highest conductance state}
    \acro{HWA}{hardware-aware}
    \acro{IMC}{in-memory computation}
    \acro{LCS}{lowest conductance state}
    \acro{LFR}{logical fault rate}
    \acro{MAC}{multiply--accumulate}
    \acro{MND}{memristive neural decoder}
    \acroindefinite{MND}{an}{a}
    \acro{MVM}{\mbox{matrix--vector} multiplication}
    \acro{NN}{neural network}
    \acroindefinite{NN}{an}{a}
    \acro{PFR}{physical fault rate}
    \acro{QEC}{quantum error correction}
    \acro{RNN}{recurrent neural network}
    \acroindefinite{RNN}{an}{a}
    \acro{ReLU}{rectified linear unit}
    \acro{SFQ}{since single-flux quantum}
    \acroindefinite{SFQ}{an}{a}
    \acro{SRAM}{static random access memory}
    \acroindefinite{SRAM}{an}{a}
    \acro{TIA}{transimpedance amplifier}
\end{acronym}


\section*{Introduction}

\Ac{FTQC} holds the promise of solving extremely
difficult problems with efficient time and space complexity~\cite
{Shor_1994, Gidney_2021}. However, this efficiency is at the cost
of resource-intensive classical procedures that are required for protecting the
logical quantum state of the quantum processor against noise~\cite
{Preskill_1998}. This includes, \emph{(a)} physically protecting the quantum state
by cooling and isolating the quantum processor, and
\emph{(b)} performing quantum control and quantum error correction protocols.
The former is the reason many quantum
technologies operate at cryogenic temperatures. But, the latter requires classical
computing circuits that have historically been designed and manufactured to
operate at room temperature and their high heat dissipations hinders their use
inside cryostats.

The transfer and processing of the data generated by \ac{QEC} protocols present
numerous challenges: \emph{(i)} With the anticipated tens of millions of physical
qubits needed for \ac{FTQC}, the repeated microsecond-long error correction cycles
would produce terabytes of syndrome data per second to be processed for
days- or even months-long computations~\cite{Gidney_2021,
Beverland_2022, Camps_2024}. Transferring this amount of data
from the cryogenic environment to classical electronics located at room
temperature quickly leads to wiring bottlenecks~\cite{Reilly_2019}.
\emph{(ii)} Transferring small signals from measurement at the quantum processor level
through multiple temperature stages to be processed by room temperature
instrumentation requires careful amplification to mitigate the increased
thermal noise at each stage. Finally, \emph{(iii)}, \iac{FTQC} implementation of
non-Clifford gates requires active error correction, that is, real-time
processing of the syndrome data by a decoder module, and application of the
resulting recovery operations at a time scale comparable to the coherence time
of the quantum system~\cite{Roffe_2019, Battistel_2023}. Classical
processors capable of operating accurately at cryogenic temperatures can
mitigate these challenges. This requires computing with extremely low heat
dissipation, as the cooling power of dilution refrigerators is limited (\num{1} to
\qty{2}{\W} at the \qty{4}{\K} stage).

\Ac{SFQ} digital electronics are natively compatible with
cryogenic temperatures, recent works~\cite{Holmes_2020, Ueno_2021,
Ueno_2022a} propose \ac{SFQ}-based architectures for decoding using various
heuristic approximations to graph-matching algorithms~\cite{
 Delfosse_2021, Wu_2022, Chamberland_2020},
while~\cite{Ueno_2022a} proposes a binarized \ac{SFQ}-based neural decoders.
Those systems are highly complex and consume a lot of energy, casting doubts on
the ability to fabricate them reliably and on their potential benefits.
The main challenge in
building scalable cryogenic fault-tolerant decoders is the need for
high-density cryogenic memory blocks for storing at least two types of
information:

\begin{enumerate}[label=(\alph*)]
 \item The program data: Even the simplest decoding heuristics require
  complex computations that rely on large look-up tables to store the logic
  and execute the algorithmic steps of the decoder~\cite{Das_2022,
  Liyanage_2024}.
 \item The input data: Fault-tolerant error
  correction requires processing multiple rounds of imperfect stabilizer
  measurements (typically as a graph-based algorithm on a 3-dimensional
  lattice) at once. Moverover, current decoding logic depends on historical
  progression of the algorithm~\cite{Tan_2023, Skoric_2023,
  Bombin_2023}. Therefore, some type of memory must store
  and recall a historical state of computation.
\end{enumerate}

In this paper, we introduce a \ac{MND} which combines
the advantages of \ac{RNN} and the low-power consumption
of resistive memory arrays to eliminate the issues related to storage and recall
for both types of memory blocks. In neural decoders, the program data (item \emph{(i)} above) is the
weights and biases of a neural network~\cite{Chamberland_2018a}. This
information is stored in the conductance states of the resistive memory devices
which are physically tuned, thereby providing a realization of \ac{IMC}~\cite{Verma_2019}.
In addition, \iac{RNN} comprises an
\emph{internal state} which is a real-valued vector (or a
higher-dimensional tensor) that encodes a latent representation of its prior
inference (the $h$ wires in Fig.~\ref{fig:QEC}). The \ac{RNN} will therefore not need
to receive the syndrome data of multiple rounds of error correction at once,
and instead stream their processing one at a time as they get generated, while
updating its internal state in each iteration (from $h_i$ to $h_{i+1}$ in
Fig.~\ref{fig:QEC}).

More specifically, we investigate TiO$_x$-based analog resistive memory devices
using TiN electrodes~\cite{Mesoudy_2022}. A crossbar arrangement of these
memory cells enables the \ac{MVM} operations
at the heart of neural network algorithms to be performed natively by relying
on Ohm's law and Kirchoff's circuit law, thus removing the time- and
energy-intensive process of moving data from memory to processing
units~\cite{Berggren_2020}. Such non-volatile memristive devices have small
footprints~\cite{Chua_1971, Strukov_2008, Chua_2011, Song_2023}, and benefit
from \ac{CMOS}-compatible fabrication processes~\cite{Mesoudy_2022}, data retention
time of up to \num{10} years~\cite{Kumar_2017}, and analog switching
dynamics~\cite{Woo_2018}, making them promising candidates for efficient \ac{MVM} in
terms of processing time, energy dissipation, and
scalability~\cite{Sebastian_2020, Amirsoleimani_2020, Hu_2016}. Furthermore,
they operate very well at cryogenic temperatures~\cite
{Beilliard_2020, Mouny_2023b, Mouny_2023a}, are robust to temperature variations,
and can be calibrated to adjust to long-term drifts in the input signal or
environmental noise.

Moreover, the \ac{MND} processes neural activations (the feature tensors propagating
forward along the network) as inherently analog signals which are never
converted to digital data in binary representation. This allows for a much
better footprints compared to the alternative \ac{CMOS}-based digital (binary)
memory technologies such as \ac{SRAM} and
resistive random access memory which were previously considered for
low-power digital neural decoding~\cite{Wang_2020, Ichikawa_2022}. We note that
the mixed-signal nature of \ac{MND} makes it attractive for soft
decoding~\cite{Ali_2024} by directly interfacing the readout resonator
and eliminating long and error-prone amplification and measurement of syndrome
bits. Since it is difficult to scale neural decoders up for large surface code
patches, we envision that such a tightly integrated decoder--QPU hybrid system
may be augmented with further (more global) decoding stages such as large
scale Union--Find or collision clustering decoders~\cite{Liyanage_2023,
Barber_2023}.

Our early \ac{MND} prototype is fabricated in \ac{CMOS} \qty{180}{\nano\metre} and demonstrates cryogenic compatibility down to \qty{35}{\K}~\cite{Mouny_2024}.
It exhibits a decoding delay of \qty{1}{\micro\second} per stabilizer measurement round, which is similar to Collision Clustering decoders able to perform real-time \ac{QEC} by demonstrating a $2\times2$ stability experiment ~\cite{Riverlane_2024}.
Our decoder delay is currently bottlenecked by the pulse width used at cryogenic temperature and could be reduced to \qty{200}{\nano\second} with minimal \ac{CMOS} design optimization.
It is expected that our processing delay will not increase significantly with larger \ac{QEC} codes as analog \ac{RNN} allows for parallel computation.
Additionally, the \ac{MND} prototype consumes \qty{3.4}{\milli\watt} at \qty{35}{\kelvin} to perform the \ac{RNN} computation and \qty{10.9}{\milli\watt} for the auxiliary electronics used to interface with the memristors.
It is anticipated that the distance-3 \ac{MND} will consume $\sim$\qty{50}{\milli\watt} for the \ac{RNN} computation and roughly \qty{120}{\milli\watt} for the memristor/\ac{CMOS} interfacing electronics in \ac{CMOS} \qty{180}{\nano\metre}.
The power consumption can be significantly reduced by using smaller \ac{CMOS} nodes, e.g., \qty{22}{\nano\metre} FDSOI exhibits power consumption up to \num{70} times smaller than \ac{CMOS} \qty{180}{\nano\metre}~\cite{Stillmaker_2017}, yielding to a \qty{2.5}{\milli\watt} power consumption per distance-3 \ac{MND}.

However, non-idealities of memrsitor arrays are known to deteriorate \acp{NN}'
performance~\cite{Joshi_2020, Wang2019, Xi_2021, Chakraborty_2020, Yon_2022}, our
study is focused on the impact of key TiO$_\textrm{x}$-based devices'
non-idealities on the accuracy of \acp{MND}. Stuck-at fault devices have the greatest
impact on the decoder's performance. Therefore, we propose and implement two
techniques for mitigating their detrimental effect. The
\ac{HWA} method consists of improving \acp{NN}' robustness by re-training it while
taking into account typical hardware constraints. In contract, the
\ac{DS} re-training method uses the exact location of stuck-at
fault devices to specifically adapt to the imperfections of a given device. We
show that the latter approach allows for high-fidelity decoding of the
distance-3 surface code.

Our paper is structured as follows. First, we provide a formal definition of
the decoding problem using a distance-3 surface code, we describe the \ac{NN}
architecture of our decoder, and present the corresponding memristive decoder
circuit. We then experimentally characterize key non-idealities of the
TiO$_\textrm{x}$-based resistive memory devices (e.g., programming variability,
stuck-at fault rate, retention time). We then assess the impact of these
non-idealities in simulations, and demonstrate that the \ac{HWA} and \ac{DS} re-training
methods recover most of the ideal neural decoder's fidelity. Finally, we discuss
the engineering advantages of leveraging analog \ac{IMC} hardware for \ac{QEC} decoding,
and provide some future perspectives.

\section*{Problem Statement}

\subsection*{Decoding the Surface Code}

\begin{figure*}
\includegraphics[width=.9\linewidth]{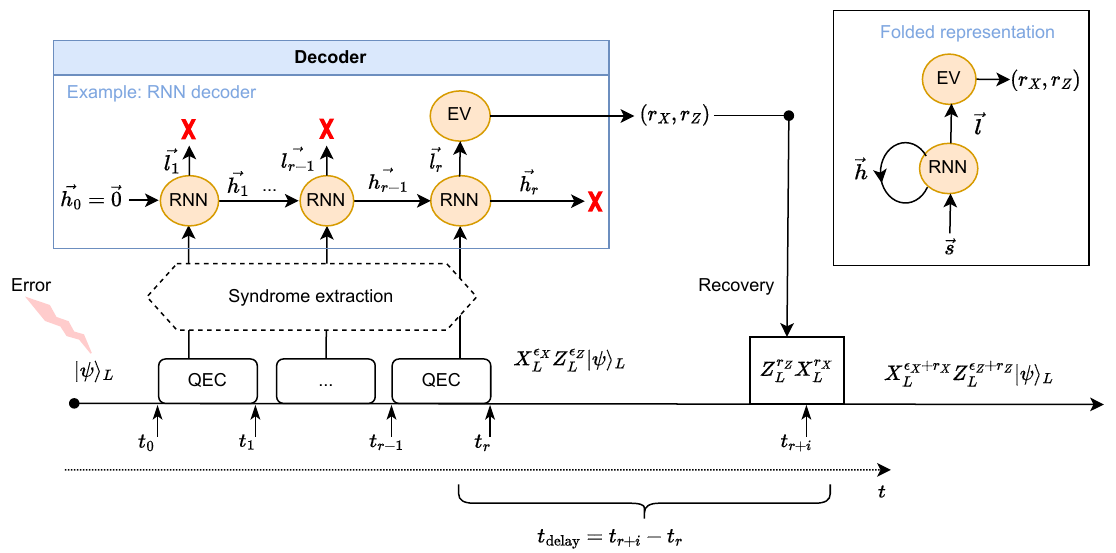}
\caption{\label{fig:QEC} \textbf{Decoding process.} In
 \acf*{QEC}, syndrome extraction rounds are performed cyclically to protect an
 encoded quantum state $\ket{\psi}_L$ from logical errors. Typically, the $X$
 and $Z$ stabilizers' syndromes are processed separately but simultaneously by
 two independent decoding modules. The syndrome measurement outcomes are fed to
 the decoder modules, which analyze them to produce a decision output
 $(r_X,r_Z)$. This decision output is used to construct the recovery operation
 $Z_{L}^{r_Z}X_{L}^{r_X}$. Between the end of the last \acs{QEC} round at $t_r$ and
 the application of the recovery operation at $t_{r+i}$, idling errors can
 affect the qubits, and they will not be taken into account by the decoding
 algorithm. The encoded state just prior to the application of recovery
 operations is therefore $X_{L}^{\epsilon_X}Z_{L}^{\epsilon_Z}\ket
 {\psi}_L$, where $\epsilon_{x,z}$ represents the cumulative effect of the
 errors that occur during the \acs{QEC} rounds and the idling time $t_{\textrm
 {delay}}$. Following the application of the recovery operation (here assumed
 to be ideal for simplicity), the final encoded logical quantum state
 is $X_{L}^{\epsilon_X+r_X}Z_{L}^{\epsilon_Z+r_Z}\ket{\psi}_L$. Here,
 the example decoder is based on a \acf*{RNN}. Each new
 syndrome is used as the input to the \acs{RNN} and the hidden state $\vec{h_i}$ is
 passed to the subsequent recurrence. At the end of the \acs{QEC} process, the output
 state $\vec{l_r}$ is passed to an evaluation module \acf*{EV} (the fully connected
 output layer in the case of \iac{RNN}) to provide the final binary output of the
 decoder. On the right-hand side of the figure, the standard folded
 representation of this \acs{RNN} is illustrated. A more detailed schematic
 architecture of this \acs{RNN} decoder is depicted in Fig.~\ref{fig:RNN}.}
\end{figure*}

In Fig.~\ref{fig:QEC}, the process of active error correction via stabilizer
measurements and decoding is schematized. Here, successful error correction
amounts to matching the decoder-proposed recovery operator $(r_X, r_Z)$ with
the logical error $(\epsilon_X,\epsilon_Z)$ afflicting the logical state
$\ket{\psi}_L$ encoded by the error correcting code. The performance
of the decoder is dependent on its speed due to idling errors associated with
the decoding delay. Therefore, minimizing the decoding time as much as possible
is desirable.

\begin{figure*}
\includegraphics[width=.9\linewidth]{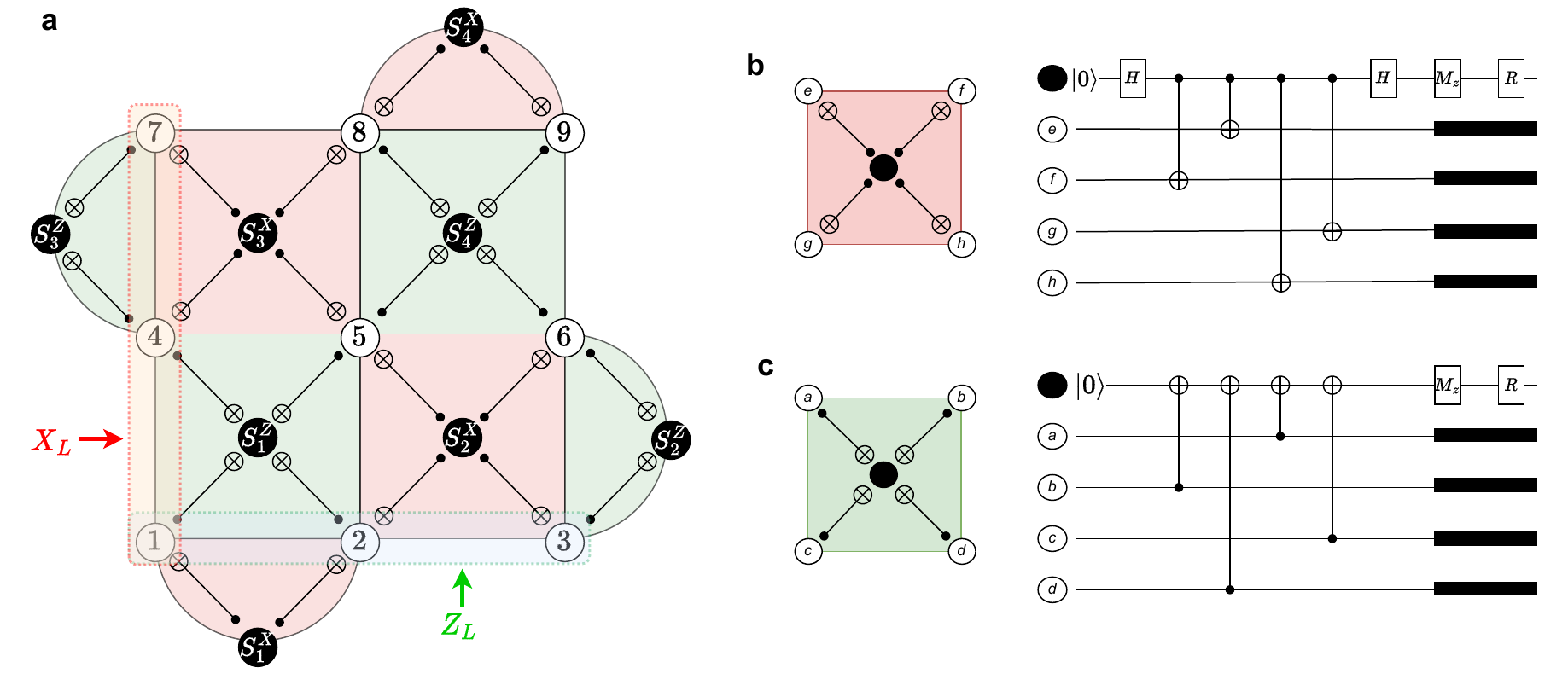}
\caption{\label{fig:surface_code} \textbf{Surface code and stabilizer
 measurements.} (a) Distance-3 rotated surface code, also called
 surface-17. Data qubits \num{1} to \num{9} are identified by circles filled in white, and
 syndrome qubits are identified by circles filled in black. Each syndrome qubit
 resides on a ``plaquette'' corresponding to a specific stabilizer measurement
 $S$. The sequence of operations realizing an $X$ stabilizer (pink plaquette)
 and a $Z$ stabilizer (green plaquette) are shown in (b) and (c). In this
 example, a logical operator $X_L$ ($Z_L$) can be realized by applying a chain
 of physical $X$ ($Z$) operators on data qubits between the top (left) and
 bottom (right) edge of the grid. (b) Sequence of circuit operations between a
 syndrome qubit (circle filled in black, the top qubit in the circuit) and its
 neighbouring data qubits, $e$, $f$, $g$, and $h$, realizing the stabilizer
 measurement $X_eX_fX_gX_h$. It consists of syndrome qubit initialization,
 Hadamard gates ($H$), \textsc{CNOT} gates, a projective measurement along the
 z-axis ($M_Z$), and a reset ($R$) to the $\ket{0}$ state, regardless of the
 measured outcome. (c) Sequence of circuit operations between a syndrome qubit
 and its neighbouring data qubits, $a$, $b$, $c$, and $d$, realizing the
 stabilizer measurement $Z_aZ_bZ_cZ_d$. It consists of syndrome qubit
 initialization, \textsc{CNOT} gates, a projective measurement along the z-axis
 ($M_Z$), and a reset ($R$) to the $\ket{0}$ state, regardless of the measured
 outcome. The black rectangles represent idling of the data qubits, which is
 important for simulation of the circuits. More details on the exact procedure
 used for the simulation of these parity-check circuits is provided in
 Supplementary Information Note~\ref{sec:note-quantum}.}
\end{figure*}

We consider the distance-3 rotated surface code, which can be realized with
\num{17} physical qubits~\cite{Bombin_2007, Horsman_2012, Tomita_2014}. As shown in
Fig.~\ref{fig:surface_code}(a), the qubits are arranged on a square lattice,
comprising data qubits (circles filled in white) and ancilla, or syndrome,
qubits (circles filled in black). Data and syndrome qubits differ only in terms
of their function within the code, and they can be implemented using physical
systems such as superconducting circuits, trapped ions, quantum dots, or
topological qubits. Each qubit interacts with its neighbours in a specified
manner. The order and mechanism of interaction is determined by the stabilizers
being measured, as shown in Fig.~\ref{fig:surface_code}(b) and~(c) for the
stabilizers $X_eX_fX_gX_h$ and $Z_aZ_bZ_cZ_d$, respectively.

\subsection*{Neural Decoder Architecture}
\label{sec:NNarchitecture}

We consider \iac{RNN} decoder module similar to the ones described in
Refs.~\cite{Chamberland_2018a} and~\cite{Baireuther_2018}.
It may be difficult to train
neural decoders for arbitrarily large topological codes; however, we note that
the largest topological patch that must be actively decoded during \ac{FTQC} depends
on the largest entangling gate between a logical magic state and other logical
qubits~\cite{Litinski_2019}. Additionally, neural decoders are great
candidates for inclusion in distributed~\cite{Varsamopoulos_2020b} and
hierarchical decoding schemes~\cite{Delfosse_2020} in order to
create larger-scale decoding systems. We restrict our benchmarking to
the $X$ syndromes because the performance would be the same for the $Z$
syndromes. Supplementary Information Note~\ref{sec:note-quantum}
describes the model used to simulate the
quantum circuit and obtain the syndromes datasets and labels for training
the \ac{RNN} decoder. The generated syndromes dataset~\cite{Yon_2024b}
is used to train and test the \ac{RNN}.

\begin{figure*}
\includegraphics[width=1.0\linewidth]{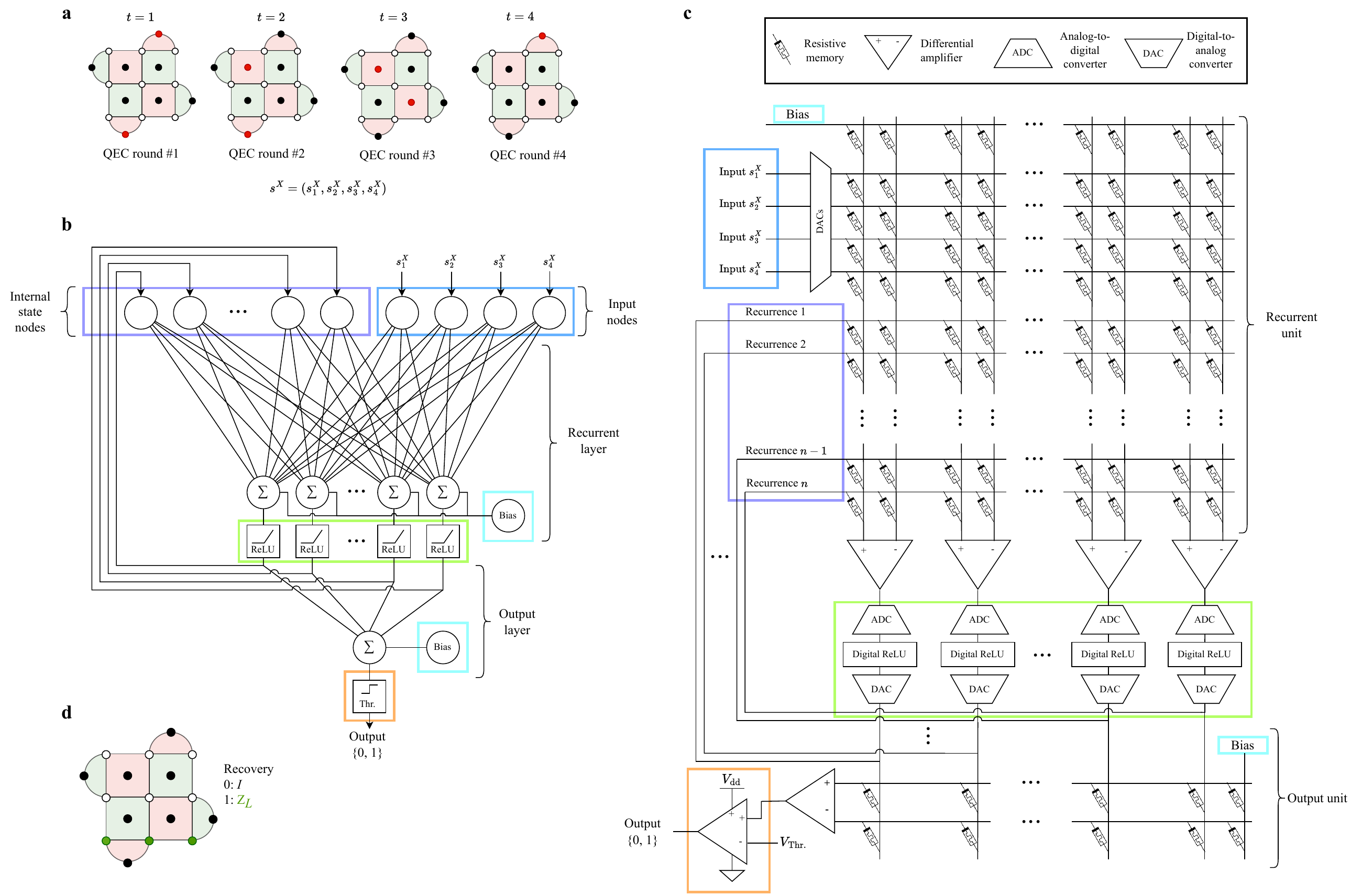}
\caption{\label{fig:RNN} \textbf{Recurrent neural network decoder architecture
 and corresponding memristive decoder circuit.} \Acf*{QEC} is
 handled as a sequential classification problem. (a) Error correction
 rounds performed by measuring stabilizer operators on the lattice of
 physical qubits. Ancilla qubits (the circles filled in red or black
 representing an error or no error being detected, respectively) measure the
 bit-flip $X$ (phase-flip $Z$) stabilizers $\hat{S}_{\{1-4\}}^X$
 ($\hat{S}_{\{1-4\}}^Z$). Measured eigenvalues of ancillas are mapped to
 classical bits to build the syndromes $s^{X}$ and $s^{Z}$, which serve as
 inputs to the neural decoder. Here, we consider only the $s^{X}$ syndromes,
 obtained at times $t=1, 2, 3, 4$. (b) Recurrent neural network decoder
 architecture for the distance-3 surface code. It comprises a recurrent
 layer and an output layer. Bias is also applied to both layers.
 The syndrome at $t=1$ is the initial input along with the initialized hidden
 state $\vec{h_0}$. The output of the recurrent layer is routed back to be used
 as new input to the internal state nodes, as a new syndrome($t=2$) is provided
 to the input nodes. This is repeated until the last \acs{QEC} round at time $t=n$
 (here, $n=4$), when the output of the recurrent layer is forwarded to the
 output layer, where the value of the single output neuron is compared to a
 threshold value to produce a binary classification. (c) Memristive decoder
 circuit. It is composed of two arrays of memristors. The first is applied
 recursively (recurrent array) until $t=n$, and the second acts as the
 classifier (output array). Input syndromes are converted from the digital
 domain to the analog domain by digital-to-analog
 converters so that the memristive crossbar can perform the analog
 matrix--vector multiplication. The memristor conductances are
 programmed according to the digitally trained weights of the equivalent
 neural network decoder. Differential amplifiers subtract the outputs from each
 pair of memristor columns to obtain the resulting node's value. This
 signal is forwarded to the activation function array (green rectangle), where
 an analog-to-digital conversion is performed to apply a digital \ac{ReLU} activation function.
 The signal is converted back to the analog domain before
 being routed back to the recurrence input ports or sent to the output
 array. Finally, the value of the output array is passed through a
 comparator to produce a binary classification. An additional memristor
 row in each array allows the application of a bias signal. (d) Recovery
 operation. To conclude the \acs{QEC} process, a recovery operation is applied to the
 data qubits according to the classification provided by the decoder. The
 identity is applied in the case where no logical error has been detected
 (output \num{0}), and the $Z_L$ operator is applied (circles filled in green) if a
 logical error has been detected (output \num{1}).}
\end{figure*}

Our \ac{RNN} architecture is illustrated in Fig.~\ref{fig:RNN}(b). It consists of
a fully connected recurrent layer and a fully connected output evaluation layer.
There are \num{4} input nodes for receiving the syndrome data of an error
correction round and \num{32} internal state nodes to store the hidden state of the
previous round via recurrent connections.
Assuming similar error rates for physical gates and measurements, the
distance-3 code requires three \ac{QEC}
cycles to be fault tolerant~\cite{Chamberland_2018b, Chamberland_2018a,
Baireuther_2018}. The outputs of the recurrent layer, after application of the
activation function, a \ac{ReLU} function in this case,
are fed back to the input of the internal state nodes when the next error
correction round's syndrome data ($s^{X}$) arrive to the input nodes, and
the process is repeated for a total of at least \num{3} cycles, as
illustrated in Fig.~\ref{fig:RNN}(a). In order to evaluate the logical error
rate of the scheme, we measure the data qubits in the final cycle
(see Supplementary Information Note~\ref{sec:note-quantum}).

After the fourth round has been provided to the input nodes, the output
of the recurrent layer is forwarded to the output layer and passed
through a threshold function. Subsequently, the neural decoder outputs a single
binary result, \num{0} or \num{1}, indicating whether a logical error has occurred at the
end of the \ac{QEC} rounds (see Fig.~\ref{fig:RNN}(d)).

\subsection*{Memristive Neural Decoder}

We present a memristive electronic circuit to implement the neural decoder
architecture discussed above, where the parameters of the neural network are
stored in crossbar arrays of TiO$_x$ resistive memory.
The \ac{MND} architecture is shown in Fig.~\ref{fig:RNN}(c). It comprises two
distinct memristor arrays: the recurrent array, which
maps the weights corresponding to the recurrent layer, and the
output array, which maps the weights corresponding to the
output layer~\cite{Gokmen_2018}, in accordance with the architecture shown
in Fig.~\ref{fig:RNN}(b). Input ports receive data from syndrome measurements in
the form of voltage pulses. Between the two crossbars of memristive devices, we
perform an analog-to-digital conversion to apply a digital \ac{ReLU} activation
function. The output is then converted back into an analog signal in the form
of a voltage pulse. Performing the activation function in the digital domain
simplifies the circuit design by removing concerns about analog signal
deterioration, as it offers full control over the shape of the signal sent into
the second layer. The range and resolution of
\ac{ADC} and \ac{DAC} is discussed in Supplementary
Information Note~\ref{sec:note-non-idealities}. Finally, the output port
provides the binary classification result of the decoder, which determines the
recovery operation to be applied to the surface code.

The memristor crossbar arrays needed to perform analog \ac{MVM} consist of two TiN
electrodes separated by a TiO$_\textrm{x}$-based switching layer~\cite{Mesoudy_2022}.
Following an initial non-reversible electroforming process, a
conductive filament containing oxygen vacancies is created within the oxide
layer~\cite{Milo_2020}, which can be subsequently at least partially dissolved
and re-established through voltage pulses applied on the electrodes, leading to
programmable, non-volatile conductance states for the memristive
device~\cite{Alibart_2012}.

It is therefore possible to map the weight matrix of a \ac{NN} layer into the
conductance states $G_{jk}^\pm$ of the  memristors inside a crossbar array. To
implement both positive and negative weights, a differential pair of memristive
devices is used (see Fig.~\ref{fig:RNN}(b)). The weight-to-conductance mapping
procedure is given by
\begin{equation}
\label{eq:2}
    G_{jk}^\pm = |w_{jk}| \frac{G_{\textrm{HCS}
    } - G_{\textrm{LCS}}}{w_{\textrm{max}}} + G_{\textrm{LCS}},
\end{equation}
where $w_{\textrm{max}}$ is the absolute maximum weight of a given layer, and
$G_{\textrm{HCS}}$ and $G_{\textrm{LCS}}$ are the \acp{HCS} and \acp{LCS}, respectively.
In other words, they are the maximum and minimum values that can be programmed
on TiO$_\textrm{x}$-based memristive devices. If $w_{jk} > 0$, $G_{jk}^+$ is
programmed with respect to Eq.~(\ref{eq:2}) and $G_{jk}^-$ is set to
$G_{\textrm{LCS}}$. If $w_{jk} < 0$, $G_{jk}^-$ is programmed with respect to
Eq.~(\ref{eq:2}) and $G_{jk}^+$ is set to $G_{\textrm{LCS}}$. After the \ac{RNN}
training has completed, the weights can be mapped to conductance states.

A crossbar configuration allows memristive devices to realize \ac{MVM} natively, by
relying on Ohm's law and Kirchhoff's current law. In Fig.~\ref{fig:RNN}(c), the
current output of each column in the crossbar array is the sum of the
input rows' voltages multiplied by the effective conductance values of the
differential pairs of memristors. From the circuit laws, we have
\begin{equation}
\label{eq:1}
    i_k = \sum_{j} \left(G_{jk}^+ - G_{jk}^-\right) v_j,
\end{equation}
where $i$ and $v$ denote output current and input voltage, respectively. For
each differential amplifier, the symbols ``$+$'' and ``$-$'' set in superscript
form denote the polarity of the pins wired to the memristors.
From Eq.~(\ref{eq:1}), each column implements naturally the \ac{MAC} operation~\cite{Ielmini_2018}.

\section*{Electrical Characterization}
\label{sec:charac}

In this section, we describe and experimentally characterize various hardware
non-idealities of \mbox{TiO$_\textrm{x}$-based} resistive memory devices. Due
to the variability of the fabrication process and switching mechanisms, these
devices exhibit multiple non-idealities~\cite{Zhang_2020, Adam_2018, Wang_2019,
Yon_2022}, such as read variability, programming variability, and stuck-at
fault malfunction (that is, the memory devices become stuck in either \ac{HCS} or \ac{LCS}
after electroforming or shortly after a conductance programming
attempt~\cite{Chen_2015a}). These non-idealities are expected to decrease the
fidelity of \iac{MND}. Other non-idealities such as random telegraphic noise and
$1/f$ noise are also commonly reported for oxide-based
devices~\cite{Yon_2022}; however, we do not to investigate their impact, as it is
expected to be insignificant for high-speed \acp{MND} operating with input voltage
pulses in the nanosecond range. Regarding conductance state retention,
Supplementary Information Fig.~\ref{fig:retention} shows no noticeable change in
the conductance state after \num{8}~hours, confirming the stability of the memory
state of these devices, acceptable for target applications. Furthermore, we have
recently demonstrated data retention at \qty{4.2}{\kelvin} for over \num{15}
minutes~\cite{Mouny_2023b}, suggesting that TiO$_\textrm{x}$-based resistive
memory devices are good candidates for a cryogenic \ac{MND}.

\begin{figure*}
\hspace{-13mm}
\includegraphics[width=.8\linewidth]{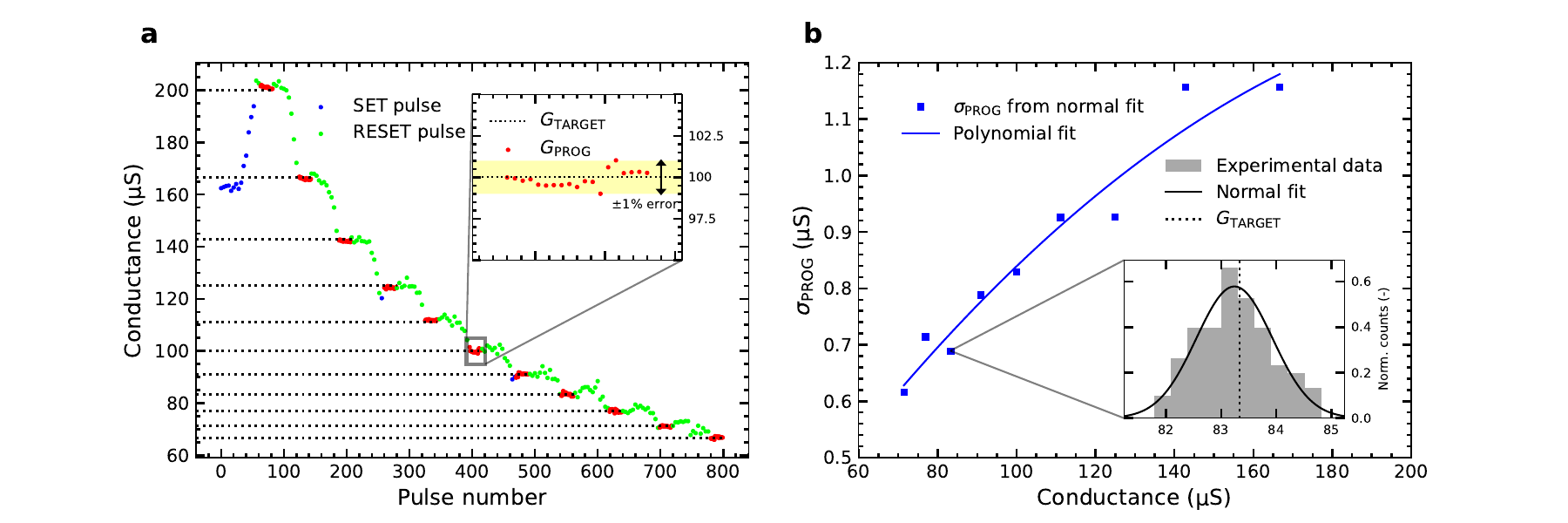}
\hspace{-14mm}
\includegraphics[width=.335\linewidth,
 trim=0 1.5mm 0 0, clip]{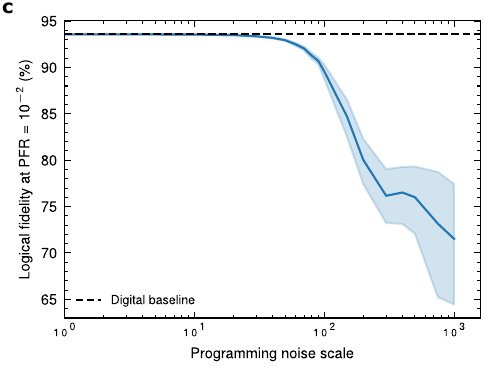}
\hspace{-4mm}
\caption{\label{fig:prog-noise} {\textbf{Programming variability
 characterizations of TiO$_\textrm{x}$-based resistive memory and its impact on
 the neural decoder.}} (a) Pulse programming of \num{11} conductance states of a
 TiO$_\textrm{x}$-based resistive memory device. Positive voltage pulses induce
 a conductance increase (labeled ``SET pulse'') due to the growth of the
 conductive filament inside the TiO$_\textrm{x}$ layer between the electrodes.
 Negative voltage pulses are employed to decrease the conductance
 (``RESET pulse'') due to the rupture of the conductive filament. The dotted
 black lines denote the target conductance (``$G_\mathrm{TARGET}$''), whereas
 the red dots denote the programmed conductance (``$G_\mathrm{PROG}$'') values
 upon convergence of the closed-loop read--write--verify algorithm. The inset
 is a zoom-in on the readout of the \qty{100}{\micro\siemens} state within the
 allowed error of \qty{1}{\percent}. (b) Programming variability model based on the
 programming standard deviations of multiple conductance states. The multilevel
 programming cycle shown in (a) is conducted \num{10} times in a double sweep for \num{10}
 devices. For each conductance state, the
 standard deviation of the distribution of programmed values is extracted to fit
 a programming variability model, except for the \acf*{HCS} and the \acf*{LCS}.
 (c) Decoding fidelity of a distance-3
 surface code for a typical physical fault rate of $10^{-2}$ using a trained
 \acf*{RNN}, as a function programming variability of the weights, represented as a
 factor of the polynomial fit used in (b). The error bands represent the \qty{95}{\percent}
 confidence interval over \num{10} random seeds. The performance of the \acs{RNN} remains
 close to the digital baseline before the noise standard deviation reaches a
 factor of $\sim$\num{10} times higher than the experimental value, after which the
 fidelity drops significantly.}
\end{figure*}

Programming variability, also called cycle-to-cycle variability, is responsible
for the inaccurate mapping of trained weights to the conductance states of
memristors. Figure~\ref{fig:prog-noise}(a) shows the programming procedure of
\num{11} conductance states on our fabricated memristors using a closed-loop
read--write--verify algorithm~\cite{Alibart_2012} (see Supplementary
Information Note~\ref{sec:note-charac}). It can be seen that the target values
given by Eq.~(\ref{eq:2}) cannot be reached exactly due to the stochastic nature
of the resistive switching process. Therefore, a programming variability model
that accounts for cycle-to-cycle variability and device-to-device variability
has been obtained from experimental characterizations of
TiO$_\textrm{x}$-based resistive memory devices~\cite{Mouny_2023b}. The
characterization process is detailed in Supplementary Information Note~\ref
{sec:note-charac}. In this model, the actual programmed conductance values
are expressed as
\begin{equation}
\label{eq:3}
    G_{jk}^\pm \leftarrow G_{jk}^\pm + \mathcal{N}(0, \sigma_{\text{prog}}(G_{jk}^\pm)),
\end{equation}
where $\sigma_{\text{prog}}(G_{jk}^\pm)$ follows the polynomial fit of
Fig.~\ref{fig:prog-noise}(b). The median value of the relative variations, that
is, $\sigma_{\text{prog}}(G_{jk}^\pm) / G_{jk}^\pm$, is \qty{0.8}{\percent} for
TiO$_\textrm{x}$-based resistive memory devices in the conductance range of
[60, 200]~\unit{\micro\siemens}.

The device yield of a chip is usually slightly under \qty{100}{\percent} due to variations
and imperfections in the nanofabrication process. This translates to a non-zero
probability of stuck-at fault devices. In this case, the device cannot be
programmed to the desired conductance for mapping a given weight of the \ac{NN}.
Recent work on passive crossbars of memristors has shown a probability of
stuck-at fault devices on the order of $\sim$\qty{1}{\percent}~\cite{Kim_2021, Jang_2022,
Yeon_2020}. In the case of our TiO$_\textrm{x}$-based resistive memory devices,
this probability can reach $\sim$\qty{10}{\percent} and the devices are usually stuck in
their \ac{HCS}.

\section*{Results}
\label{sec:results}

Due to the hardware non-idealities characterized in the previous section, it is
expected that the transfer of a digitally trained neural decoder to the
equivalent memristor-based implementation would decrease its performance. We
simulate the impact of these key non-idealities on the fidelity of \iac{MND}
implemented in a mixed-signal \ac{IMC} architecture. We then benchmark mitigation
strategies using re-training allowing for almost complete compensation for the
negative impact of the non-idealities of memristors.

We first evaluate the impact of the programming variability observed in
TiO$_\textrm{x}$-based memristive devices by investigating their effects on the
decoder performance. Figure~\ref{fig:prog-noise}(c) shows the evolution of the
decoding fidelity of a distance-3 \ac{RNN} for a typical physical fault rate of
$10^{-2}$ when the programming noise is increased. The experiment is repeated
\num{10} times to obtain a mean fidelity value (shown using a blue curve). Note that
no noticeable fidelity decrease is observed below about \num{10} times the
experimental noise value, therefore, the programming variability of
TiO\textbf{$_\textrm{x}$}-based resistive memory devices does not represent a
roadblock for the realization of the \ac{MND}. Evaluating the impact of the
resolution of the \acp{DAC} and \acp{ADC} (see Supplementary Information
Fig.~\ref{fig:dac_adc_res}) leads to the same conclusion, as an \num{8}-bit
discretization of the input and output values does not present any significant
impact on the performance of the decoding. Lastly, the neural decoder appears
to be robust against the reading variability (up to \qty{1}{\percent}) induced by the analog
electronics during \iac{MVM} operation (see Fig.~\ref{fig:output_noise}). The
methodology used to evaluate the impact of the \acp{DAC} and \acp{ADC}, and the reading
variability is described in Supplementary Information
Note~\ref{sec:note-non-idealities}.

\begin{figure*}
\centering

\begin{subfigure}[b]{0.46\textwidth}
  \raisebox{16 em}[\height][0pt]{\llap{\textbf{a\quad}}}
  \raisebox{6 em}{\includegraphics[width=\textwidth]{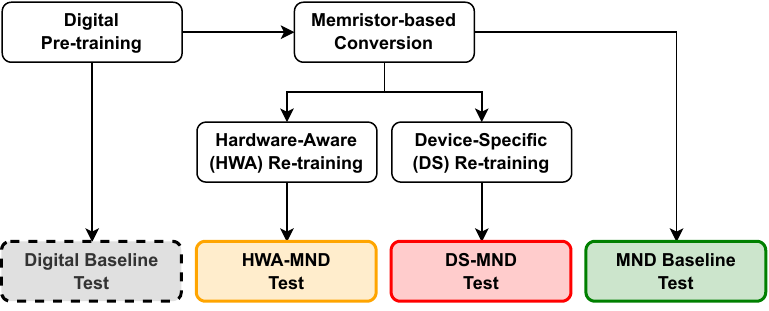}} 
  \label{fig:methods}
\end{subfigure}
\hfill 
\begin{subfigure}[b]{0.46\textwidth}
  \raisebox{18 em}[\height][0pt]{\llap{\textbf{b\quad}}}
  \includegraphics[width=\textwidth]{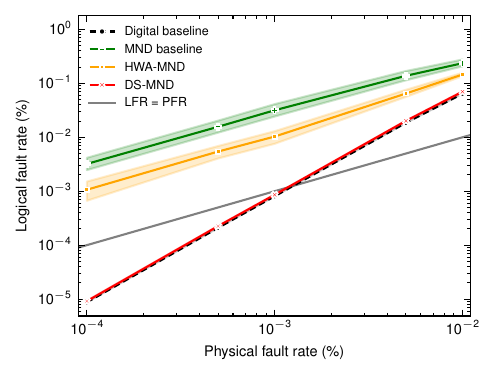}
  \label{fig:pfv_vs_lfr}
\end{subfigure}

\begin{subfigure}[b]{0.46\textwidth}
  \raisebox{18 em}[\height][0pt]{\llap{\textbf{c\quad}}}
  \includegraphics[width=\textwidth]{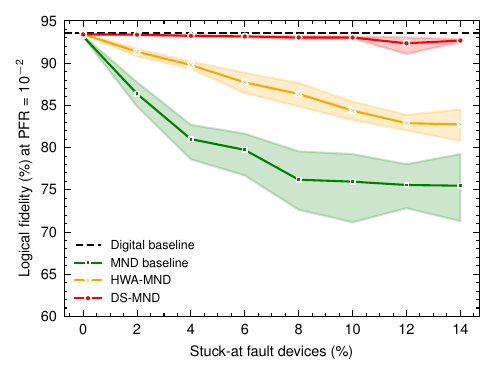}
  \label{fig:scheme_stuck_evolution}
\end{subfigure}
\hfill
\begin{subfigure}[b]{0.46\textwidth}
  \raisebox{18 em}[\height][0pt]{\llap{\textbf{d\quad}}}
  \includegraphics[width=\textwidth]{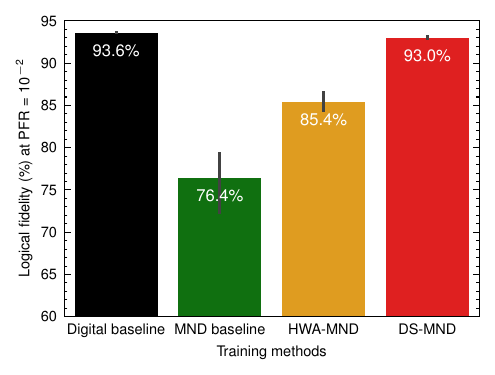}
  \label{fig:training_schemes}
\end{subfigure}

\caption{\label{fig:mnd-results}
\textbf{Baselines and re-training methods for the memristive neural decoder.}
(a) Schematic representation of the methods used to train the \acf*{RNN} decoder. The
fidelity of the digital baseline is tested immediately after classical
training. The \acf*{HWA} and \acf*{DS} \acf*{MND} methods benefit from a re-training step after being converted to
a memristor-based model with the \emph{IBM Analog Hardware Acceleration
Kit}~\cite{Rasch_2021}, whereas the \acs{MND} baseline training method is converted and
tested without any re-training. (b) \acf*{LFR} after decoding as
a function of the \acf*{PFR}. The intersection between the line
``\acs{LFR} = \ac{PFR}'' and decoder performances represent the breakeven points
(i.e., the pseudo-threshold values) of the corresponding neural decoders.
(c) Logical fidelity percentage of the decoder at a \acs{PFR} of $10^{-2}$, for the
baselines and the re-trained decoders, as a function of the percentage of
stuck-at fault devices. For the \acs{HWA}-\acs{MND}, the stuck-at fault devices define the
dropconnect rate used during re-training. (d) Logical fidelity percentage of
the decoder at a \acs{PFR} of $10^{-2}$ for the baselines and the re-trained decoders.
For the results presented in (b) and (d), the rate of stuck-at fault devices is
set at the expected fabrication yield of \qty{10}{\percent}. For the results presented in
(b), (c), and (d), the digital baseline fidelity is obtained with an equivalent
neural decoder. The error bars and shaded regions represent the \qty{95}{\percent} confidence
interval over \num{10} random seeds.}
\end{figure*}

We then study the impact of stuck-at fault devices on the fidelity of the neural
decoder, which is expected to reach to up to \qty{10}{\percent} for the evaluated memristor
technology. To emulate the impact of this non-ideality, a random subset of
network parameters is set to zero when testing the decoder. This situation
corresponds to the case where one of the memristor of a differential pair is
stuck in \ac{HCS} or \ac{LCS} and the other is either also stuck in the same state or
purposefully programmed in order to bring the logical analog weight to zero.
The results presented in Fig.~\ref{fig:mnd-results}(c) (green curve) show that,
without any re-training designed to tackle this non-ideality, the stuck-at
fault rate significantly reduces the performance of the decoder. A decrease of
more than \qty{17}{\percent} in fidelity is observed when the percentage of stuck-at fault
devices during inference reaches \qty{8}{\percent} (this is equivalent to \qty{15}{\percent} of the
model parameters, since a parameter is encoded via a pair of devices).

Therefore, we explore two re-training methods to attempt to restore the initial
fidelity of the \ac{MND} (see Fig.~\ref{fig:mnd-results}(a)), which we compare to
two baselines:

\begin{itemize}
\item Digital baseline: an ideal version of the neural decoder that can perform
 syndrome processing at room temperature on classical hardware. The parameters
 are trained using a deep learning method~\cite{Lecun_2015} (see Supplementary
 Information Table~\ref{table:meta-parameters}) and are represented as 32-bit
 floating-point variables that do not include hardware non-idealities. The \ac{RNN}
 architecture is based on Ref.~\cite{Chamberland_2018a}. The digital baseline
 performance is represented in black in Fig.~\ref{fig:mnd-results}, which we
 use as an upper bound and a reference against which to compare the other
 methods.

\item \Ac{MND} baseline: a na\"ive implementation of an analog memristor-based neural
 decoder that could be integrated in a cryogenic environment. The \ac{RNN}'s
 architecture is identical to the digital baseline, but the parameters are
 converted into the equivalent memristor conductance values after digital
 training. Hardware constraints and memristor non-idealities are simulated
 based on experimental characterizations (see Supplementary Information
 Table~\ref{tab:simulation-parameters}), but no re-training is used. The \ac{MND}
 baseline performance is represented in green in
 Fig.~\ref{fig:mnd-results}, which is expected to be a lower bound for other
 \ac{MND} methods.

\item Hardware-aware memristive neural decoder (\ac{HWA}-\ac{MND}): the \ac{MND} baseline
 augmented with a re-training post-processing. The re-training consists of one
 additional training epoch during which different random weights are set to \num{0}
 at each forward pass. The dropconnect method~\cite{Wan_2013} is typically used
 for regularization, but in the context of \ac{HWA} re-training it is incorporated
 to improve the robustness of the \ac{RNN} against stuck-at fault devices. For this
 reason, we set the dropconnect rate to match the expected number of weights
 blocked due to stuck-at fault devices in the characterized hardware
 (see Supplementary Information Fig.~\ref{fig:drop_connect} for the different
 stuck-at fault rates). At the testing stage, the hardware non-idealities are
 simulated (see Supplementary Information
 Table~\ref{tab:simulation-parameters}). The performance of \ac{HWA}-\ac{MND} is
 represented in yellow in Fig.~\ref{fig:mnd-results}.

\item  Device-specific memristive neural decoder method (\ac{DS}-\ac{MND}): a variation of
 the \mbox{\ac{HWA}-MND}, where the dropconnect is not random but follows a specific
 device. During the entire re-training epoch, the weights corresponding to the
 stuck-at fault memristors in a given crossbar are fixed at \num{0}. This approach
 forces the \ac{RNN} to precisely adapt its parameters to this hardware limitation.
 The performance of \ac{DS}-\ac{MND} is represented in red in
 Fig.~\ref{fig:mnd-results}.
\end{itemize}

Supplementary Information Note~\ref{sec:note-simulation} provides details about
the implementation of training and inference in our experiments, and we have
made the simulation data available in Ref.~\cite{Yon_2024a}.

Dropping connections randomly during the \ac{HWA}-\ac{MND} re-training provides a generic
mitigation strategy applicable to any crossbar of memristors given that the
average number of stuck-at fault devices is known prior to re-training.
However, it does not allow for a full recovery of the digital baseline
performance even for low rates of stuck-at fault devices. \ac{DS}-\ac{MND} addresses this
shortcoming. This re-training of \ac{DS}-\ac{MND} completely avoids updating weights that
cannot be reliably programmed in the crossbar. The idea of this \ac{DS} re-training
is to leverage precise knowledge of the crossbars' electrical characterizations
to find the best \ac{NN} parameters for that particular memristive circuit. It has
the advantage of converging towards a nearly optimal solution, at the cost of
having to individually characterize each crossbar of the \ac{MND} to localize
stuck-at fault devices, a task which can be performed through methods such as
march tests~\cite{Goor_1993, Chen_2015b}. The simulation results, with the
expected \qty{10}{\percent} rate of stuck-at fault devices(see Fig.~\ref{fig:mnd-results}
(d)), show that only \ac{DS}-\ac{MND} reaches within \qty{1}{\percent} of the digital baseline's
fidelity, whereas the \ac{HWA}-\ac{MND} recovers only about half of the performance loss
due to hardware non-idealities.

Fig.~\ref{fig:mnd-results}(b) represents the decoding performance of the \acp{MND}
studied in the case of different \acp{PFR} for a fixed
percentage of stuck-at fault devices of \qty{10}{\percent} to match the typical fabrication
yield. Only the \ac{DS} re-training method maintains the pseudo-threshold of the
memristive decoder near the pseudo-threshold of the baseline decoder.
Therefore, based on the characterized non-idealities of TiO$_\textrm
{x}$-based resistive memory devices, it is a necessity to introduce specific
knowledge of the chip during re-training to achieve the highest decoding
performance in the case of a distance-3 code based on the \ac{RNN} we have studied.
The general solution provided by \ac{HWA}-\ac{MND} provides robustness against
device-specific hardware issues, but remains insufficient.

\section*{Discussion}
\label{sec:discussion}

In this section, we discuss our novel hardware approach for implementing \iac{RNN} decoder relying on the \ac{IMC} paradigm to perform \ac{MVM} on passive crossbars of memristors. We implemented a simulation based on the \emph{IBM Analog Hardware Acceleration Kit}~\cite{Rasch_2021} that accounts for the experimentally characterized non-idealities of \mbox{TiO$_\textrm{x}$-based} resistive memory
devices and measured their impact on our neural decoder's performance. By
applying computational methods to mitigate key hardware non-idealities, we
improved the robustness of the neural decoder. In particular, we found that
using the dropconnect method during re-training can greatly improve the fidelity of \iac{MND} that whose performance is significantly reduced by stuck-at fault devices.
Moreover, we have seen that localizing stuck-at fault memory devices in the crossbars and using that knowledge to disable the corresponding connections during re-training can lead to a fidelity score comparable to that of the digital baseline (i.e., having a \qty{<1}{\percent} difference).
Furthermore, the \ac{MND} exhibits only a small drop in the pseudo-threshold for the distance-3 surface code in numerical simulations (see Fig.~\ref{fig:mnd-results}(b)).
Therefore, we can conclude that our results support the effectiveness of specialized training methods for \acp{MND} to achieve near-optimal performance.

Although an experimental proof of concept has not yet been performed on fully integrated hardware, the results we have presented offer a promising
pathway to realizing high-fidelity neural decoders using \ac{IMC} and analog
memristive devices.
One interesting avenue for future research is the development of a
fully analog version of the memristive decoder circuit presented in this paper. Such
an implementation would bring further benefits in terms of the decoding time and energy efficiency. Analog activation functions
have been reported in the literature~\cite{Krestinskaya_2020}.
For instance, the \ac{ReLU} function can be
applied in the analog domain using multimodal transistors~\cite{Isin_2022}. It is
therefore possible to envision \acp{MND} using an analog activation function after the first layer instead of implementing \iac{ADC} followed by a digital activation function followed by a \ac{DAC}, as presented in this paper. A practical circuit would also
necessitate many more electronic components to realistically perform the
decoding task, some of which we did not consider here. For example,
\acp{TIA} would be needed to convert a current signal to a voltage signal
at the output of each memristor row. Also, an analog memory unit might be necessary to store and transmit the hidden states' signals back to the recurrence input ports.

In our work, we made the assumption that imperfections arising from analog or
digital \ac{CMOS} components (e.g., noise introduced by differential amplifiers, \acp{TIA}, and
\acp{ADC}/\acp{DAC}; divergence of the analog activation function from its digital
counterpart in the case of a fully analog implementation; and signal distortion arising
from multiple recurrences) are much less critical in terms of the fidelity of the \ac{MND} in comparison to the non-idealities exhibited by the
memristors. However, future work should include a more exhaustive
analysis of the impact of circuit-level imperfections (e.g., using training methodologies that have been introduced~\cite{Liu_2014, Liu_2015}).

Scalability is a well-known issue for \ac{QEC} decoders: as the error correction code distance increases, the syndrome space grows
exponentially. Neural decoders thus require an exponentially larger dataset to
learn about correlations between syndromes and the occurrences of logical errors.
Our decoder cannot overcome the issue of scalability. However, it can be
integrated in a multi-stage or hierarchical decoding scheme~\cite{Varsamopoulos_2020b, Delfosse_2020, Meinerz_2022}.
Multi-stage or hierarchical approaches have been proposed to improve the efficiency of \ac{NN}-based
decoders by using a combination of two decoding modules; in some instances~\cite{Varsamopoulos_2020a}, the first module is a simple classical decoder, and the second is a \ac{NN}-based decoder. The role of the \ac{NN}-based module is to act as a supervisor, identifying when the correction
suggested by the simple decoder will lead to a logical error. This approach
results in a constant execution time once the \ac{NN} has been trained,
regardless of the physical error rate, and scales linearly with the number of
qubits in the code. A more recent study~\cite{Chamberland_2022} demonstrates
that \acp{NN} can be used as local decoders to remove an initial set of
errors, thereby reducing the syndrome space and enabling the fast execution of
a global decoder (minimum-weight perfect matching in the study) to correct the remaining errors. In this sense, a fully analog version of our \ac{MND} could be
integrated in a hierarchical decoding strategy and act as a local decoder to
feed inputs to a global decoder.

Another challenge related to the scalability of decoders is the cryogenic compatibility of
the chosen hardware. Indeed, as the number of physical qubits increases, to
avoid a wiring bottleneck between the control electronics at room temperature
and the quantum processor in a cryogenic environment, it is highly beneficial to
integrate the decoder hardware directly within the cryostat~\cite{Reilly_2019}. Within this scope, the energy consumption and thus the heat dissipation of the decoder should be minimized to avoid perturbations in the quantum system. 
The \ac{MND} presented in this paper shows promise in terms of cryogenic compatibility as the \ac{MAC} operations rely on an energy-efficient memristive \ac{IMC} architecture instead of digital circuit blocks such as multipliers and adders. Even if current hardware
implementations of \acp{NN} employing \acp{ASIC} and \acp{FPGA} are not optimized in
comparison to CPU-based approaches, they continue to suffer from the delays and
energy expenditure associated with digital \ac{MAC} operations. Furthermore, in recently proposed approaches introducing the idea of quantized \ac{IMC} for \ac{QEC}~\cite{Wang_2020, Ichikawa_2022}, the implementations still
require digital multipliers and adders.
From their fast inference time and energy efficiency, \acp{MND} are a promising technology for direct integration in a dilution refrigerator.
However, the cryogenic compatibility of a memristor-based fully analog integrated circuit for \ac{QEC} and the detailed characterization of its decoding time and power dissipation remains to be investigated.

\section*{Code and data availability}

The code from our study is available from the corresponding author upon
reasonable request. The syndromes dataset has been made publicly
available~\cite{Yon_2024b} as well as the model training and simulation
output data~\cite{Yon_2024a}.

\section*{Acknowledgements}

We thank our editor, Marko Bucyk, for his careful review and editing of the
manuscript.
The authors acknowledge the financial support received through the NSF’s CIM
Expeditions award (CCF-1918549).
P.~R.~acknowledges the financial support of Mike and Ophelia Lazaridis,
Innovation, Science and Economic Development Canada (ISED), and the Perimeter
Institute for Theoretical Physics.
Research at the Perimeter Institute is supported in part by the Government of
Canada through ISED and by the Province of Ontario through the Ministry of
Colleges and Universities.
This work was supported by the Natural Sciences and Engineering Research Council
of Canada (NSERC).
LN2 is a French-Canadian joint International Research Laboratory
(IRL-3463) funded and co-operated by the Centre National de la Recherche
Scientifique (CNRS), the Universit\'e de Sherbrooke, the Universit\'e de Grenoble
Alpes (UGA), the \'ecole Centrale de Lyon (ECL), and  the Institut National des
Sciences Appliqu\'ees de Lyon (INSA Lyon). It is supported by the Fonds de
Recherche du Qu\'ebec -- Nature et Technologie (FRQNT).

\section*{Author contributions}

All authors contributed to this article and approved of the submitted version.

Victor Yon: methodology, software implementation, run simulations, results analysis and visualization, writing--original draft preparation.

Frédéric Marcotte: methodology, software implementation, run simulations, hardware characterization, writing--original draft preparation.

Pierre-Antoine Mouny: methodology, software implementation, hardware characterization, manuscript editing.

Gebremedhin A. Dagnew: methodology, syndrome dataset generation, manuscript editing.

Bohdan Kulchytskyy: syndrome dataset generation, manuscript review, supervision.

Sophie Rochette: methodology, manuscript editing, supervision.

Yann Beilliard: methodology, manuscript editing, supervision.

Dominique Drouin: manuscript review, supervision.

Pooya Ronagh: methodology, manuscript editing, supervision.

\section*{Competing interests}

The authors declare no competing interests.

\clearpage
\bibliography{main}

\clearpage
\onecolumngrid

\setcounter{section}{0}
\setcounter{equation}{0}
\setcounter{figure}{0}
\setcounter{table}{0}
\setcounter{page}{1}

\makeatletter

\renewcommand{\theequation}{S\arabic{equation}}
\renewcommand{\thefigure}{S\arabic{figure}}
\renewcommand{\thetable}{S\arabic{table}}
\renewcommand{\thesection}{S\arabic{section}}
\renewcommand{\bibnumfmt}[1]{[#1]}
\renewcommand{\citenumfont}[1]{#1}
\acresetall 

\begin{center}
\textbf{\large Supplementary Information:\\
A Cryogenic Memristive Neural Decoder for Fault-Tolerant Quantum Error Correction}
\end{center}

\section{Simulation of Quantum Stabilizer Circuits}
\label{sec:note-quantum}

Here we provide some background on the simulation of surface codes and the error
model used to generate the dataset for the neural decoder's training and
testing (the dataset is available for download~\cite{Yon_2024b}). Our
experiment is based on the distance-3 rotated surface code. This surface code
corresponds to a 17-qubit system in which there are \num{9}~data qubits
(circles filled in white in Fig.~\ref{fig:surface_code}(a) of the main
manuscript) and \num{8} ancilla qubits (circles filled in black) corresponding to the
\num{8} stabilizer generators $S_1^X,S_2^X,S_3^X,S_4^X,S_1^Z,S_2^Z,S_3^Z$, and
$S_4^Z$.

In our study, we rely on Stim~\cite{Gidney_2021} to generate the circuits shown in Fig.~\ref{fig:surface_code} required to simulate the rotated surface code, PyMatching~\cite{Higgott_2021} for decoding with minimum-weight perfect matching in order to benchmark the relative performance of the \ac{RNN} decoder, and a custom-built glue code to integrate Stim and PyMatching. While writing this paper, the need for the glue code was eliminated due to recent updates in PyMatching2.

 We simulate the memory-$X$ rotated surface code, where preparation and final measurements are done in the $X$-basis. In addition, we conduct this simulation under circuit level noise, which means the following: 
\begin{itemize}
    \item with a noise probability of $p$, each two-qubit gate is followed by a two-qubit Pauli error drawn uniformly and independently from $\{I, X, Y, Z\}^{\otimes 2}\setminus\{I \otimes I\}$;
    \item with a probability of $2p/3$, the preparation of the $|0\rangle$ state is replaced by the state $|1 \rangle = X|0\rangle $. Similarly, with a probability of $2p/3$, the preparation of the $|+\rangle$ state is replaced by the state $|-\rangle = Z|+\rangle$;
    \item with a probability of $2p/3$, any single-qubit measurement has its outcome flipped; and
    \item with a probability of $p$, each idling-qubit location is followed by a Pauli error drawn uniformly and independently from the stabilizer set $\{X, Y, Z\}$. 
\end{itemize}

A Stim circuit is first initialized with a probability $p$ for a given code distance and a number of rounds. This circuit is then repeatedly called to generate samples. The steps taken during circuit generation for performing the parity-check rounds shown in Fig.~\ref{fig:surface_code}(b) and (c) of the main manuscript are detailed as follows:

\begin{enumerate}
    \item At the onset of the \ac{QEC} cycle, the data qubits are initialized in the $|+\rangle$ state (for memory-$X$ experiments), while ancilla qubits are set to the $|0\rangle$ state. These are replaced with the states $|+\rangle$ and $|1\rangle$, respectively, with a probability of $2p/3$, accounting for preparation errors.
    \item Before the beginning of every parity check round, each data qubit is depolarized with a probability of $p$ due to idling.
    \item To perform the parity-check circuits depicted in Fig.~\ref{fig:surface_code}(b) and (c), Hadamard gates are applied to the \mbox{$X$-syndrome qubits,} converting the $|0\rangle$ states to $|+\rangle$ states and the $|1\rangle$ states to $|-\rangle$ states, followed by single-qubit depolarizing noise with a probability of $p$.
    \item Next, four CNOT cycles are executed, and each cycle is succeeded by two-qubit depolarizing noise with a probability of $p$, following the configuration outlined in Fig.~\ref{fig:surface_code}(b) and (c).
    \item Hadamard gates are applied to the $X$-syndrome ancilla qubits, followed by depolarizing noise with a probability of $p$.
    \item With a measurement error probability of $2p/3$, all ancilla qubits are measured in the $Z$-basis, after which they are reset to the $|0\rangle$ state.
    \item To simulate preparation errors for the next \ac{QEC} round, bit flips are applied with a probability of $2p/3$ on the ancilla qubits, along with idling, and steps \num{2} to \num{7} are repeated to conduct repeated syndrome measurement cycles.
    \item Finally, the data qubits are measured in the $X$-basis with a measurement error a probability of $2p/3$. The data qubit measured implicitly define one final round of computational-basis-type stabilizers (in this case $X$ stabilizers) inferred via the corresponding parity-check matrix. Since this is a classical process and is not subject to the circuit-level noise defined during a typical stabilizer round, the syndromes extracted in this round are also referred to as ``perfect syndromes''. The final state of the encoded logical qubit is then determined. 
\end{enumerate}

The first stabilizer measurement round is the encoding round, which creates an entangled logical qubit. In the absence of an error, the encoding results in a logical $|+\rangle$ state since the data qubits are initialized in a logical $|+\rangle$ state. Measuring a $|-\rangle$ logical state in the end thus indicates a logical error, and the goal of a decoder is to predict such events.   

To generate samples for training the \ac{RNN}, we first separate the $X$ and $Z$ syndromes, focusing only on the $X$ syndromes. Next, we divide the $X$-syndrome measurements into separate time steps, corresponding to each syndrome extraction round. The training label represents the logical bit-flip value, where \num{0} indicates that the logical state corresponding to the given syndromes has remained unchanged, and 1 indicates that the logical state has flipped. The training, validation, and testing samples are generated using noise probabilities between $p = 0.00001$ and $p = 0.01$, along with 3 rounds of stabilizer measurement and then a reset.

A key preprocessing step before feeding measurements to the decoder for analysis is to take the differences between consecutive measurement outcomes. Syndromes indicate the presence or absence, as well as the type, of errors that have occurred to the physical qubits. The \ac{RNN} outputs a prediction of $0$ or $1$, representing the absence or presence of a logical error, respectively. The corresponding recovery operator is thus given by $(Z_L)^y$, where $y$ is the binary output of the \ac{RNN}.

\section{Other Hardware Non-idealities }
\label{sec:note-non-idealities}

\subsection*{Analog-to-Digital and Digital-to-Analog Converter Range and Resolution}

\begin{figure*}[t]
\includegraphics[width=.5\linewidth]{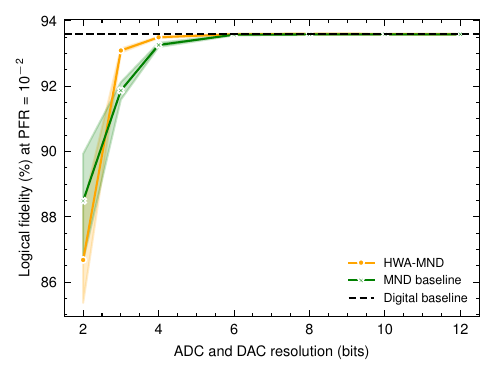}
\caption{\label{fig:dac_adc_res} \textbf{Impact of the \acf*{ADC} and \acf*{DAC} resolution on the \acf*{RNN} decoder's test fidelity.} The \acf*{PFR} is set to 10$^{-2}$. The test fidelity is obtained without including
 reading variability or defective devices, in order to assess the impact of
 resolution alone. At an \num{8}-bit resolution, that is, \num{256} quantization steps
 between the boundary values \num{-6} and \num{6}, the fidelity approaches the digital baseline fidelity (using a \num{32}-bit floating point accuracy). The error range is obtained based on \num{10} independent training rounds.}
\end{figure*}

To simulate the behaviour of \acp{ADC} and \acp{DAC} during inference on the crossbar arrays of memristors, a rounding function is applied,
\begin{equation}
    \label{eq:s11}
    f(x) =
    \begin{cases}
      -b & x < -b \\
      \text{round}\left(\frac{x}{2b} n\right) \frac{2b}{n} & -b\leq x\leq b \\
      b & x > b
   \end{cases},
\end{equation}
where $x$ represents the \ac{ADC} and \ac{DAC} input, its range is given by $[-b,b]$, and $n$ is the
number of quantization steps, for example, \num{256} for \num{8}-bit resolution. The
round function rounds the floating-point value to the closest integer.
Based on the typical \ac{MVM} output values, we fix the \ac{ADC} range to
 $b=6$. The \ac{DAC} range is fixed to $b=1$ to match with the sigmoid output range. In \acp{ADC} and \acp{DAC}, there is a trade-off between resolution and acquisition frequency,
that is, a lower resolution yields a higher acquisition frequency. An
application where inference time needs to be minimized, such as with the neural
decoder, requires a high acquisition frequency. From Fig.~\ref{fig:dac_adc_res}, we can see that an \num{8}-bit resolution is appropriate to reach the
highest acquisition frequency without substantially degrading the \ac{RNN} decoder's
fidelity.
Therefore, we simulate \ac{ADC} and \ac{DAC} discretization during the testing of a memristive neural decoder (\ac{MND}), but no re-training strategy seems necessary to maintain the decoder's fidelity.

\subsection*{Reading Variability}
Every analog electronic component induces a small reading variability.
Analog-to-digital converters and digital-to-analog converters~\cite{Spear_2023}; differential amplifiers; and memristors~\cite{Wang_2018} involved in \ac{MVM} operations are no exception.
Based on experimental characterizations~\cite{Mouny_2023b}, we estimate that the dominant source of reading variability should come from memristors.
To reduce the computational complexity, we simplify the simulation by considering every source of reading variability as one Gaussian random value added at the output of each row of memristors. 
As a conservative estimate, we chose a Gaussian distribution with a standard deviation equal to \qty{1}{\percent} of the maximum logical value (defined as $b$ in the previous section).

\begin{figure}
    \centering
    \includegraphics[width=0.5\linewidth]{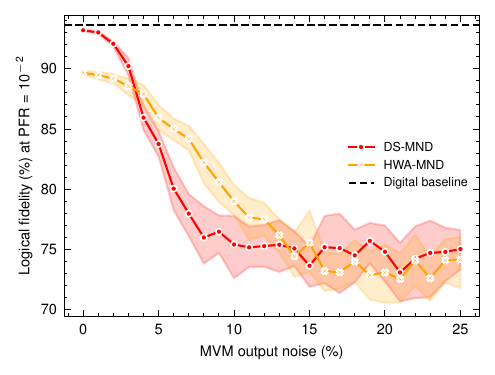}
    \caption{\textbf{Impact of the \acf*{MVM} output noise on the \acf*{RNN} decoder's test fidelity.} The \acf*{PFR} is set to 10$^{-2}$. A random Gaussian value is added to the output of each memristor row to account for the expected reading noise produced by the electronic components and the memristors. The percentage is relative to the maximum logical output value (set to \num{6} in our simulations). All other non-idealities are kept as defined in Table~\ref{tab:simulation-parameters}.}
    \label{fig:output_noise}
\end{figure}

The effect of different reading noise on the neural decoder's fidelity is illustrated in Fig.~\ref{fig:output_noise}.
This simulation result suggests that a \qty{1}{\percent} reading noise is not detrimental to our decoding application.
However the performance drops quickly when the value increases above this limit.
We can observe that the random dropconnect applied with our hardware-aware memristive neural decoder (\ac{HWA}-\ac{MND}) re-training method also increases the robustness of the model against high reading variability.
The device-specific memristive neural decoder (\ac{DS}-\ac{MND}) re-training method is more sensitive to perturbations that are not related to the stuck-at fault devices; this is visible by the faster fidelity drop in Fig.~\ref{fig:output_noise} when the output noise is increased in comparison to the \ac{HWA}-\ac{MND} method.
This reading variability study shows that realistic noise values do not impact the performance enough to require a re-training method addressing specifically this issue.
However, it is worth noting that a dropconnect strategy, applied in addition to the \ac{DS}-\ac{MND} method, could mitigate the negative effect of reading variability.

\section{Experiments on TiO$_\textrm{x}$-Based Resistive Memory Devices}
\label{sec:note-charac}

The studied TiO$_\textrm{x}$-based resistive memory devices are characterized by a Keysight Technologies
B1500A semiconductor device parameter analyzer that has a \num{200}~million samples/second
waveform generator / fast measurement unit (WGFMU) using a Lake Shore CPX-VF probe
station at room temperature. For all measurements, the electrodes at the bottom are
grounded and the signals are applied to the electrodes at the top.

\subsection*{Conductance State Programming}

In out study, the resistance of memristors is programmed using a closed-loop
read--write--verify algorithm~\cite{Alibart_2012, Mouny_2023b}. This algorithm
allows the programming of a memristor to a target resistance by
applying successive read and write pulses. Initially, a device's resistance is
read using a read pulse (having a \qty{0.2}{\V} amplitude and a \qty{1}{\micro\second} pulse
width). If this measured resistance is larger than the target within a \qty{1}{\percent}
tolerance, a positive write pulse is applied ($\sim$1~V amplitude and \qty{200}{\nano\second}
pulse width); if it is lower, a negative pulse is applied. A read pulse is then
applied to check the new resistance state, which starts a new cycle. The pulse
amplitude is increased after each consecutive pulse of the same polarity by a
constant step in order to converge to the target resistance. Using this
read--write--verify algorithm for various target levels of resistance, we can
perform multilevel programming as shown in Fig.~\ref{fig:prog-noise}(a) in the main manuscript.

\subsection*{Programming Variability Characterization}

The programming variability model is based on experimental measurement conducted on
TiO$_\textrm{x}$-based resistive memory devices. As shown in Fig.~\ref{fig:prog-noise}(a) in the main manuscript, multilevel
programming is performed from the high-conductance states to the low-conductance
states. To assess the cycle-to-cycle programming variability (shown in
Fig.~\ref{fig:prog-noise}(b)), we successively perform multilevel programming from the low-conductance states to the high-conductance
states \num{10} times. For each
state and at each multilevel programming pass, the conductance is read \num{20} times
and the mean value is kept. This process is repeated for \num{10} memristors
to include device-to-device variability in the programming error model. The
histogram of the mean programmed values for a conductance state is fitted by a
Gaussian distribution and its standard deviation is extracted (also depicted in
Fig.~\ref{fig:prog-noise}(b)). The fit of the standard deviation $\sigma$ corresponds to the programming variability which depends on the conductance of the programmed state.

\subsection*{Conductance State Retention}
\begin{figure}[b]
\includegraphics[width=.9\linewidth]{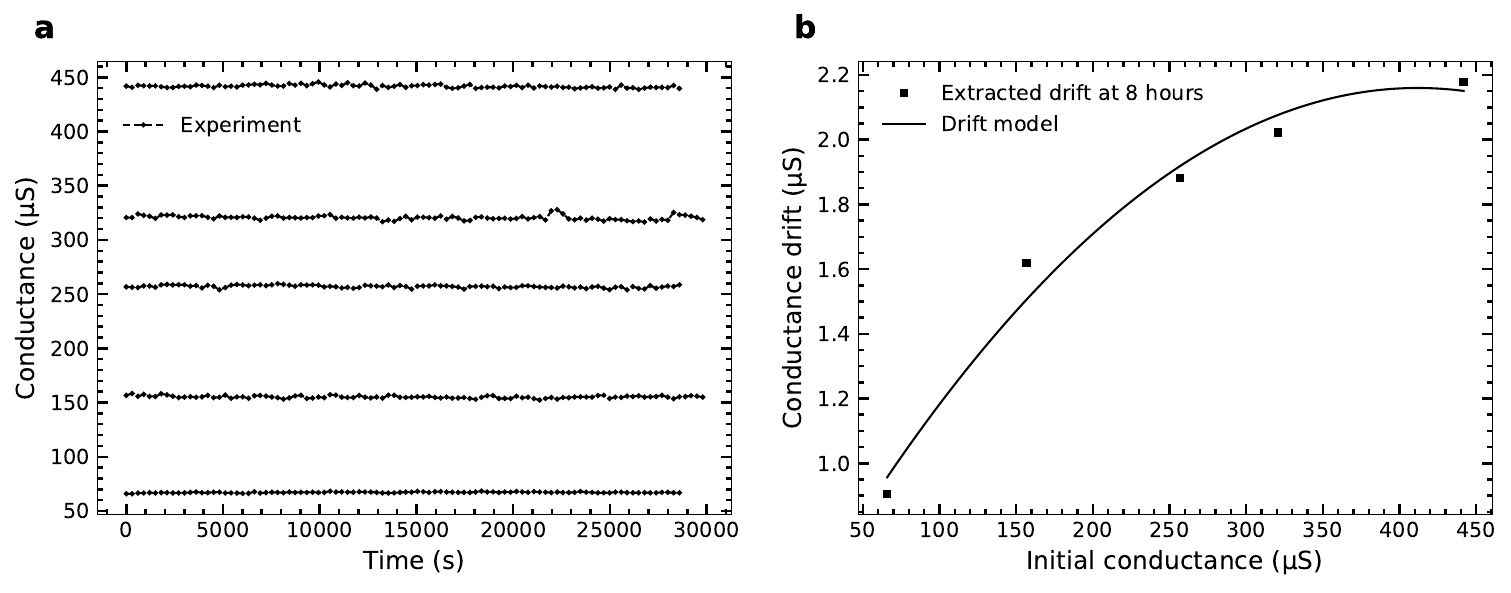}
\caption{\label{fig:retention} \textbf{TiO$_\textrm{x}$-based resistive memory device retention study.}
 (a) Retention of \num{5} conductance states between \qty{65}{\micro\siemens} and \qty{440}{\micro\siemens}. No conductance drift is noticeable. (b) Conductance drift of the \num{5} conductance states after eight-hour-retention measurements. The drift with respect to the initial conductance is fitted by a second-order polynomial function.}
\end{figure}

The retention measurements presented in Fig.~\ref{fig:retention} are performed at room temperature for \num{8} hours for \num{5} conductance states. After the programming of a conductance state, a \qty{1}{\milli\second}-wide pulse with a \qty{200}{\milli\volt} amplitude is applied every 5 minutes to measure the retention. The retention of the conductance states is measured successively. The 5 states demonstrate non-volatility over the eight-hour retention test. The conductance drift after 8 hours with respect to the initial conductance state can be fitted by a second-order polynomial function. Considering the small drift for this memristor technology, we have not considered this behaviour in the simulations presented in this paper.

\section{Neural network training methods}
\label{sec:note-simulation}

\subsection*{Implementation of Training and Inference}

The training of the \ac{RNN} always begins with a classical deep learning~\cite{Lecun_2015} optimization process, using PyTorch~\cite{Paszke_2019} with no consideration for hardware non-idealities (see Fig.~\ref{fig:mnd-results}(a) in the main manuscript).
The metaparameters used during this step are given in Table~\ref{table:meta-parameters}. In the case of hardware-aware memristive neural decoder (\ac{HWA}-\ac{MND}) and device-specific memristive neural decoder (\ac{DS}-\ac{MND}) re-training, the \ac{RNN}
parameters (weights and biases) are loaded from an earlier digital training, and
one additional training epoch is run to adapt the model to newly introduced hardware
constraints. Every other metaparameter remains unchanged during re-training.

\begin{table}[h!]
\begin{center}
\begin{tabular}{| r|l |} \hline 
    
    \textbf{Metaparameters} & \textbf{Values}  \\ \hline 
    
    Input size & \num{4} rounds of \num{4} binary syndromes\\
    Hidden layer size & \num{32}\\
    Activation function & \acs{ReLU}  \\
    Batch size & \num{16}\\
    Loss & Cross-entropy  \\ 
    Optimizer & Adam~\cite{Kingma_2017}  \\
    Learning rate & \num{0.001}  \\
    Training epochs & \num{4} (+1 for \ac{HWA} and \ac{DS} methods)\\ \hline
\end{tabular}
\end{center}
\caption{\textbf{Model and training metaparameters.}}
\label{table:meta-parameters}
\end{table}

The \ac{RNN} re-training and inference simulations are performed using the IBM Analog
Hardware Acceleration Kit~\cite{Rasch_2021} v0.9.0. This Python library is used to simulate
the computations executed on a crossbar array of memristors and the relevant peripheral circuit at the behavioural level. We extend the original
toolkit to consider the specific TiO$_\textrm{x}$-based resistive memory devices used in our
work. The non-idealities we discuss in the hardware characterization section of the main manuscript
are implemented in the simulator. For instance, the statistical programming
variability is calibrated on our TiO$_\textrm{x}$-based resistive memory devices.

\subsection*{Re-training methods}
\begin{table}
    \centering
    \begin{tabular}{|r|l|} \hline 
         \makecell[r]{\textbf{Hardware constraints}\\ \textbf{and non-idealities}}& \textbf{Simulated values}\\ \hline 
         Weight clipping& \num{2.5} $\sigma$ of the Gaussian distribution of each layer's weights' values\\ \hline
         \acs{DAC} resolution& \num{8} bits\\ \hline
         \acs{ADC} resolution& \num{8} bits\\ \hline
         \acs{DAC} max logical value&1\\ \hline
         \acs{ADC} max logical value&6\\ \hline
         \acs{MVM} output noise&  \makecell[l]{Gaussian noise added with a standard deviation equal to \qty{1}{\percent}\\of the maximum logical value (6)}\\ \hline
         Max memristor resistance& \num{15000} $\Omega$\\ \hline
 Min memristor resistance & \num{5000} $\Omega$\\\hline
 Memristor stuck-at-fault rate & \qty{10}{\percent}\\\hline
 Programming noise's polynomial fit& $y= - x^3(2.40 \times 10^{-5}) + x^2(1.15 \times 10^{-2}) -x(7.07 \times 10^{-2})      $\\\hline
    \end{tabular}
    \caption{Default simulated values of the hardware constrains and non-idealities in our study. These values are used to simulate the memristor-based \acf*{RNN} with the IBM Analog Hardware Acceleration Kit~\cite{Rasch_2021}.}
    \label{tab:simulation-parameters}
\end{table}

For hardware-aware re-training, the main manuscript focuses on introducing a
dropconnect training method during the re-training of the \ac{MND} to mitigate
the effect of defective devices in the chip. However, there are other
training methods that could improve performance. One such method consists in the
addition of noise on weights during forward passes of training steps to account for the programming
variability of the characterized TiO$_\textrm{x}$-based resistive memory devices. It has been shown that setting the training
weight variability approximately equal to the inference weight
variability is optimal~\cite{Joshi_2020}. However, in our case, we observed training instability issues with this method, which led to an unexpected gradient explosion in a fraction of the trained \acp{RNN}.
Therefore, this method did note benefit the  overall performance.

The study of various non-idealities highlighted that only stuck-at fault devices have a critical
effect on the performance of decoding. We attribute this to the fact that
TiO$_\textrm{x}$-based resistive memory devices exhibit a low programming variability
(in comparison to phase-change memory devices reported in~\cite{Joshi_2020}); therefore,
injecting noise during training does not significantly improve fidelity. Also,
\iac{ADC} and \ac{DAC} of 8-bit resolution is sufficient for maintaining a fidelity close to the digital baseline, as
explained in Supplementary Information Note~\ref{sec:note-non-idealities}. Thus, the input and output
quantization during training does not allow for greater fidelity.
Moreover, it is likely that the binary output of the model helps mitigate most noise-related issues since only a perturbation near the threshold value would impact the final binary output of \acp{RNN}.

Another limitation of memristors is their limited resistance range, which constrains the parameters' values.
This limitation can easily be tackled to improve test accuracy by using a method called weight clipping to trim the parameters' values in a specific range during re-training~\cite{Joshi_2020}. It allows a certain degree of control over the
weight distribution to make the programming of weights to hardware easier. It
is performed by clipping the weights of a layer, after each backward pass in the training steps, within
the range $[-\alpha \sigma; \alpha \sigma]$, where $\alpha$ represents the scale of the clipping, and $\sigma$ is the standard deviation of the weight
distribution of that layer. Based on our empirical tests, the exact value of $\sigma$ does not appear to strongly affect the decoder's performance. Thus, we decided to fix $\sigma=2.5$ for all \ac{MND} test simulations.

\begin{figure}
\includegraphics[width=.5\linewidth]{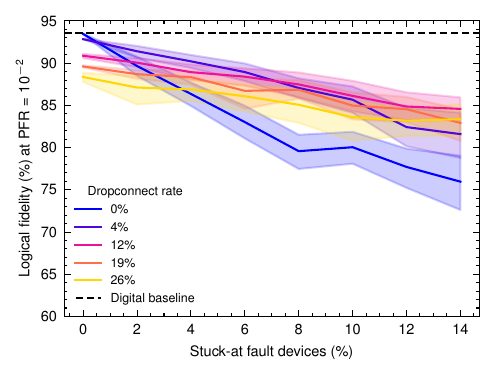}
\caption{\label{fig:drop_connect} \textbf{Impact of the dropconnect re-training method for various stuck-at-fault rates.}
Decoding fidelity at a \acf*{PFR} of $10^{-2}$, for different dropconnect probabilities during \acf*{HWA} \acf*{MND} re-training, as a function of the percentage of stuck-at fault devices during inference. The shaded error regions represent the \qty{95}{\percent} confidence interval over \num{10} random seeds.}
\end{figure}

The decoding fidelity globally improves when the dropconnect method is introduced during re-training
compared to the case where no dropconnect is employed (shown using a blue curve with the label ``\qty{0}{\percent}'' in Fig.~\ref{fig:drop_connect}).
It also appears that \iac{RNN} decoder trained with a high dropconnect rate will have a lower fidelity if no device is stuck, but its performance will decrease more slowly when the rate of stuck-at fault devices increases.
Therefore, a good trade-off between fidelity and robustness seems to be a dropconnect rate equal to the number of blocked parameters (\qty{19}{\percent}) given the expected stuck-at fault device percentage (\qty{10}{\percent}).
Since a parameter is encoded by two devices (for positive and negative values), the probability of a parameter being blocked (i.e., set to \qty{0}{\percent}) is equal to the probability of at least one of the pair of devices being stuck.

\end{document}